\documentclass[useAMS,usenatbib]{mn2e}
\usepackage{epsfig}
\usepackage{amsmath}

\newcommand{\be}{\begin{equation}}
\newcommand{\beq}{\begin{equation}}
\newcommand{\ba}{\begin{eqnarray}}
\newcommand{\ee}{\end{equation}}
\newcommand{\eeq}{\end{equation}}
\newcommand{\ea}{\end{eqnarray}}

\newcommand{\apj}{ApJ}
\newcommand{\apjl}{ApJL}
\newcommand{\mnras}{MNRAS}

\def\lsim{~\rlap{$<$}{\lower 1.0ex\hbox{$\sim$}}}

\def\gsim{~\rlap{$>$}{\lower 1.0ex\hbox{$\sim$}}}

\title[21cm emission after reionization]{The 21cm Power Spectrum After
Reionization}

\author[Wyithe \& Loeb]{J. Stuart B. Wyithe$^1$ and Abraham Loeb$^2$\\$^1$
School of Physics, University of Melbourne, Parkville, Victoria,
Australia\\$^2$ Harvard-Smithsonian Center for Astrophysics, 60 Garden St.,
Cambridge, MA 02138\\Email: swyithe@unimelb.edu.au,
loeb@cfa.harvard.edu}

\begin{document}


\maketitle

\label{firstpage}
\begin{abstract}

  \noindent We discuss the 21cm power spectrum (PS) following the
  completion of reionization. In contrast to the reionization era,
  this PS is proportional to the PS of mass density fluctuations, with
  only a small modulation due to fluctuations in the ionization field
  on scales larger than the mean-free-path of ionizing photons. We
  derive the form of this modulation, and demonstrate that its effect
  on the 21cm PS will be smaller than 1\% for physically plausible
  models of damped Ly$\alpha$ systems. In contrast to the 21cm PS observed prior
  to reionization, in which HII regions dominate the ionization
  structure, the simplicity of the 21cm PS after reionization will
  enhance its utility as a cosmological probe by removing the need to
  separate the PS into physical and astrophysical components. As a
  demonstration, we consider the Alcock-Paczynski test and show that
  the next generation of low-frequency arrays could measure the
  angular distortion of the PS at the percent level for $z\sim3-5$.

\end{abstract}


\section{Introduction}

Recently, there has been much interest in the feasibility of mapping
the three-dimensional distribution of cosmic hydrogen through its
resonant spin-flip transition at a rest-frame wavelength of
21cm~(Furlanetto, Oh \& Briggs~2007; Barkana \& Loeb~2007). Several
experiments are currently being constructed (including
MWA~\footnote{http://www.haystack.mit.edu/ast/arrays/mwa/},
LOFAR~\footnote{http://www.lofar.org/}, PAPER
~\footnote{http://astro.berkeley.edu/~dbacker/EoR/},
21CMA~\footnote{http://web.phys.cmu.edu/~past/}, GMRT~\footnote{Pen et al.~(2008)}) and more ambitious
designs are being planned
(SKA~\footnote{http://www.skatelescope.org/}).

One driver for mapping the 21cm emission is the possibility of measuring
cosmological parameters from the shape of the underlying power spectrum
(PS; see Loeb \& Wyithe 2008). During the epoch of reionization, the PS of
21cm brightness fluctuations is shaped mainly by the topology of ionized
regions, rather than by the PS of matter density fluctuations which is the
quantity of cosmological interest~(McQuinn et al.~2006; Santos et al.~2007;
Iliev et al.~2007). As a result, the line-of-sight anisotropy of the 21cm
PS due to peculiar velocities must be used to separate measurements of the
density PS from the unknown details of the astrophysics (Barkana \&
Loeb~2005; McQuinn et al.~2006). The situation is expected to be simpler
both prior to the formation of the first galaxies (at redshifts $z\ga 20$,
Loeb \& Zaldarriaga~2004; Lewis \& Challinor~2007; Pritchard \& Loeb~2008),
and following reionization of the intergalactic medium (IGM; $1\la z\la6$)
-- when only dense pockets of self-shielded hydrogen, such as damped
Ly$\alpha$ absorbers (DLA) and Lyman-limit systems (LLS) survive ~(Wyithe
\& Loeb~2008; Chang et al.~2007; Pritchard \& Loeb~2008).

In this paper we focus our discussion on the post-reionization epoch
(Wyithe \& Loeb~2007; Chang et al.~2007).  The DLAs which contain most
of the neutral hydrogen mass in the Universe at $z\la 6$ are expected
to be hosted by galactic mass dark matter halos~(Wolfe, Gawiser \&
Prochaska~2005). A survey of 21cm intensity fluctuations after
reionization would measure the modulation of the cumulative 21cm
emission from a large number of galaxies~(Wyithe \& Loeb~2008; Wyithe,
Loeb \& Geil~2008; Chang et al.~2007). Regarding the measurement of
the 21cm PS, this lack of identification of individual galaxies is an
advantage, since by not imposing a minimum threshold for detection,
such a survey collects all the available signal. This point is
discussed in Pen et al.~(2008), where the technique is also
demonstrated via measurement of the cross-correlation of galaxies with
unresolved 21cm emission in the local Universe.

Studying the 21cm PS after (rather than during) reionization offers
two advantages.  First, it is less contaminated by the Galactic
synchrotron foreground, whose brightness temperature scales with
redshift as $(1+z)^{2.6}$ ~(Furlanetto, Oh \& Briggs~2006). Second,
because the UV radiation field is nearly uniform after reionization,
it should not imprint any large-scale features on the 21cm PS that
would mimic the cosmological signatures.  In addition, on large
spatial scales the 21cm sources are expected to have a linear bias
analogous to that inferred from galaxy redshift surveys.

Most previous studies of post-reionization 21cm fluctuations have
assumed that the 21cm emission traces perturbations in the matter
density, and have considered peculiar motions only after a spherical
average~(Wyithe \& Loeb~2008; Pritchard \& Loeb~2008, Loeb \&
Wyithe~2008).  The exception is a recent paper discussing measurements
of the dark energy equation of state (Chang et al.~2007). Since the
neutral gas resides within collapsed dark matter halos, galaxy bias
plays an important role in setting the 21cm fluctuation amplitude.  In
this paper we derive the 21cm PS after reionization in the context of
the formalism that has been developed to calculate it during
reionization~(Barkana \& Loeb~2005; McQuinn et al.~2006; Mao et
al.~2008). By framing the derivation this way, the relative merits of
cosmological constraints from 21cm surveys at redshifts before and
after reionization can be more easily understood. In addition, this
formalism provides a framework to describe the possible effect of
fluctuations in the ionizing background. We then compute the
Alcock-Paczynski effect~(Alcock \& Paczynski~1979) as an example for
the cosmological utility of the post-reionization 21cm PS. In our
numerical examples, we adopt the standard set of cosmological
parameters ~(Komatsu et al.~2008), with values of $\Omega_{\rm
b}=0.24$, $\Omega_{\rm m}=0.04$ and $\Omega_Q=0.76$ for the
matter, baryon, and dark energy fractional density respectively, and
$h=0.73$, for the dimensionless Hubble constant.

\section{21cm Power Spectrum After Reionization}

The 21cm PS after reionization is expected to be dominated by the
neutral content of galaxies. In a scenario where fluctuations in the
ionizing background can be ignored, the form of the PS could therefore
be inferred directly from the galaxy PS (Loeb \& Wyithe
2008). However, we also include here the possible influence of a
fluctuating ionizing background on the 21cm PS.  Throughout our
discussion, we assume the baryonic overdensity $\delta$ to equal the
overdensity in dark-matter on sufficiently large scales. 

\subsection{The 21cm brightness temperature}
The 21cm
brightness temperature fluctuation in a region of IGM is
\begin{equation}
\label{Tb}
\Delta T=23.8\left(\frac{1+z}{10}\right)^\frac{1}{2}\left[1-\bar{x}_i(1+\delta_x)\right]\left(1+\delta\right)\left(1-\delta_v\right)\,\mbox{mK},
\end{equation}
where $\delta_v=\partial v_r/\partial r(Ha)^{-1}$ and $\partial
v_r/\partial r$ is the gradient of the peculiar velocity along the
line-of-sight.  The quantities $\bar{x}_i$ and $\delta_x$ are the mean
ionization fraction and the fluctuation in ionization fraction,
respectively. We have assumed that the spin temperature of hydrogen is much
higher than the CMB temperature [as expected for collisionally coupled gas
in collapsed objects, and observed in some DLAs~(Curran et al.~2007)]. Thus
we may neglect fluctuations in the kinetic and spin temperatures of the
neutral hydrogen gas throughout the post reionization epoch.
To evaluate the brightness temperature fluctuation we need to compute
$\delta_x$. With this goal in mind, we first write the fluctuation in
neutral hydrogen fraction as
\begin{equation}
\delta_{\rm HI}\equiv x_{\rm HI}/\bar{x}_{\rm HI}-1,
\end{equation}
where $\bar{x}_{\rm HI}$ and $x_{\rm HI}$ are the cosmic mean and local
values for the neutral fraction (mass averaged). 

\subsection{Effect of the ionizing background} 

Next we compute the effect of the ionizing background on the neutral
fraction.  Two regimes must be considered.

\subsubsection{Systems which are optically-thin  to ionizing radiation}

Low density regions of the Universe contain optically thin, Ly$\alpha$
absorbers. The ionization fraction in this regime is controlled by the
balance between the ionization rate owing to the UV background and the
recombination rate at the local gas density.  Given an ionization rate
$\Gamma$, and a hydrogen density $n_{\rm H}$ with a neutral fraction
$x_{\rm HI}$, the equilibrium in the optically thin regime is given by
the condition
\begin{equation}
x_{\rm HI}\sim\frac{n_{\rm H}\alpha_\beta}{\Gamma},
\end{equation}
where $\alpha_{\rm B}$ is the case-B recombination coefficient (Osterbrock \& Ferland~2006).
Assuming that the low density regions of the hydrogen density field
are unbiased with respect to the underlying dark matter density, we
have
\begin{equation}
x_{\rm HI}\propto (1+\delta - \frac{\Delta\Gamma}{\Gamma}),
\end{equation}
where $\Delta \Gamma$ is the perturbation in the
intergalactic ionizing background. Note that the neutral fraction in
optically thin regions is increased in overdense regions by
recombinations, and decreased by the presence of an excess ionizing
background.

\subsubsection{Optically-thick absorbers}

Since optically-thick systems are self-shielded, the effect of the UV
background on their ionization state depends on the gas distribution
within them.  Below we discuss the effect of an ionizing background on
the HI content of self-shielded systems.  Since most of the hydrogen
in such systems is known to be contained within DLAs, we focus our
discussion on DLAs.

The nature of DLAs is not understood, although they are thought to be
formed by dense, self-shielded gas in galaxies (Wolfe, Gawiser \&
Prochaska~2005).  The detailed modeling of DLAs is very uncertain,
requiring numerical simulations that go beyond the scope of this
paper. However we can utilise results of simulations in the literature
to estimate the range of possible strength for the influence of the UV
background. The column density distribution of DLAs has a power-law
form with a feature at $\sim10^{20}$cm$^{-2}$ (e.g. Storrie-Lombardi
\& Wolfe~2000). This feature has been interpreted by a number of
authors as being due to the effects of self shielding of DLAs in a
meta-galactic ionizing background (e.g. Corbelli, Salpeter \&
Bandiera~2001; Zheng \& Mirada-Escude~2002).

Zheng \& Mirada-Escude~(2002) considered the column density
distribution of damped Ly$\alpha$ systems modeled as spherical
isothermal spheres. They solved for the self shielded HI density
profile numerically and found that the neutral fraction reaches
$\sim10^{-3}$ at the radius where the system becomes optically thick
to meta-galactic ionizing radiation. Their results demonstrate that
the transition from highly ionized to highly neutral gas occurs over a
narrow region which is very small compared with the size of the
absorbing system. Moreover, they find that the radius $r_{\rm ss}$ at
which the gas becomes self-shielding is proportional to
$\Gamma^{-1/3}$. For a spherical isothermal sphere, the hydrogen mass
enclosed within radius $r$ scales as
\begin{equation}
M_{\rm H}(<r)\propto r,
\end{equation}
and so the HI mass scales as
\begin{equation}
M_{\rm HI}=M_{\rm H}(<r_{\rm ss})\propto r_{\rm ss}\propto \Gamma^{-1/3}.
\end{equation}
We therefore find that a given perturbation in the ionizing
background produces a corresponding perturbation in the mass of
neutral hydrogen within the self-shielded system of the form
\begin{equation}
\label{fluc2}
\frac{\Delta M_{\rm HI}}{M_{\rm HI}} =
-\frac{1}{3}\frac{\Delta\Gamma}{\Gamma}.
\end{equation}

Alternatively, the distribution of gas in a DLA may be better
represented by a self-gravitating disk. The effect of a meta-galactic
background on the ionization structure and starformation has been
investigated by a number of authors (Corbelli, Salpeter \&
Bandiera~2001; Schaye~2004; Susa~2008). The hydrogen column depth at
which the gas becomes self-shielding was shown to be proportional to
$\Gamma^{1/3}$ as in the spherical isothermal case
(Susa~2008). However the exponential profile of the disk models
implies that the ionized portion of the gas contains much less mass
than in the case of a more extended power-law isothermal
profile. Corbelli, Salpeter \& Bandiera~(2001) compute the column
depth of HI as a fraction of the column depth of hydrogen, at
different intensities of ionizing background. For a disk with a column
density of $10^{21}$cm$^{-2}$, the fluctuation in HI mass owing to a
fluctuation in the ionizing background is approximately
\begin{equation}
\frac{\Delta M_{\rm HI}}{M_{\rm HI}} \sim - 2\times
10^{-3}\frac{\Delta\Gamma}{\Gamma}.
\end{equation}
This effect is two orders of magnitude smaller than in the case of a
spherical isothermal sphere.

To evaluate the fluctuation in neutral fraction associated with DLAs
we assume that the neutral gas is hosted within halos of mass $M$ with
associated galaxy bias $b$. We have\footnote{The factor of
($1+\delta$) in the denominator is present because this is the
fluctuation in neutral hydrogen fraction rather than in neutral
hydrogen mass.}
\begin{eqnarray}
x_{\rm HI}&\propto& \frac{ (1+b\delta)(1-C{\Delta\Gamma}/{\Gamma})}{1+\delta}\\
&\sim& 1 + b\delta -\delta -C\frac{\Delta\Gamma}{\Gamma},
\end{eqnarray}
where the last equality has reduced the expression to lowest order in
$\delta$.

\subsubsection{Combined neutral content of the IGM}

\label{thick}
The above estimates of mass fluctuation suggest that the neutral
hydrogen content of the IGM is modified owing to fluctuations in the ionizing background ($\delta_J\equiv{\Delta\Gamma}/{\Gamma}$) by a factor of the form
\begin{equation}
\label{fluctuation}
s\approx (1-C\delta_J),
\end{equation}
where $C$ is a constant describing the magnitude of the effect. For
optically thin regions $C=1$, while for optically thick absorbers $C$
can take a range of values.  In the case of an isothermal sphere, a
fluctuation in the ionizing background leads to a fluctuation in the
HI mass of a self shielded system that is of comparable magnitude,
indicating that the value of $C$ in equation~(\ref{fluctuation}) would
be of order unity in this case. However, in halos with virial
temperatures larger than $\sim10^{4}$K, hydrogen with an isothermal
profile would be collisionally ionized, and hence have cooled into a
much more concentrated profile. Since the host masses of DLAs are
known to be greater than $10^{10}$M$_\odot$ from clustering studies
(Cook et al.~2006), an isothermal profile does not provide a
physically plausible model. On the other hand, modeling DLAs using an
exponential disk, as is appropriate for gas rich galaxies implies a
much smaller value value of $C\sim 10^{-3}$--$10^{-2}$. Thus, the
effect of fluctuations in the ionizing background on the mass averaged HI density is expected to be small ($\la1\%$).

We can now add the effects of fluctuations in ionizing background on the neutral content of optically thin and optically thick absorbers within a region of large scale overdensity $\delta$. We have
\begin{eqnarray}
\nonumber
x_{\rm HI}&=& F_{\rm thin} \bar{x}_{\rm HI} (1+\delta - \delta_J) + F_{\rm thick} \bar{x}_{\rm HI} (1+b\delta-\delta-C\delta_{J})\\
\nonumber
&&\hspace{-8mm}=\bar{x}_{\rm HI} \left[1+(F_{\rm thin} + F_{\rm thick}(b-1))\delta - (F_{\rm thin} + C F_{\rm thick})\delta_{J}\right]\\
&&\hspace{-8mm}\sim \bar{x}_{\rm HI} \left[1+(b-1)\delta - C\delta_{J}\right],
\end{eqnarray}
where $F_{\rm thin}$ and $F_{\rm thick}$ are the fractions of HI in
the optically thin and optically thick regimes respectively. In the
last equality, we assume the case where the optically thin component
contains a negligible amount of the cosmic HI. Note that we ignore the
optically thin component of the hydrogen in the remainder of this
paper. However we point out that the inclusion of this additional
component changes only the values of the co-efficients of $\delta$ and $\delta_J$ in the above
equations, but not the form of the expression.

\subsubsection{Scale dependence of fluctuations in the ionizing background}

Before deriving the fluctuation in neutral fraction we next need to
calculate the dependence of the parameter $\delta_J$ on
scale. Ionizing radiation in the IGM has a finite mean-free-path which
increases with time following the end of reionization. On scales
larger than the mean-free-path ($\lambda_{\rm mfp}$), with associated
wavenumber $k<k_{\rm mfp}=2\pi/\lambda_{\rm mfp}$, all ionizing
photons within a fluctuation were produced by galaxies that were also
within that fluctuation. In this case fluctuations in the ionizing
background simply trace fluctuations in the density of galaxies so
that the value of $\delta_J\propto b\delta$ is independent of
scale. However, on scales smaller than the mean-free-path, ionizing
photons were not produced by galaxies local to the fluctuation. As a
result $\delta_J/\delta$ is scale dependent. 

To derive this dependence we convolve the real space density field with a filter function to account for the effects of finite mean-free-path and the inverse square dependence of ionizing flux. The fluctuation in flux at a position $\vec{x}$ becomes 
\begin{equation}
\delta_J(\vec{x}) \propto \int d\vec{x}' G(\vec{x},\vec{x}') \delta(\vec{x}').
\end{equation}
For simplicity we assume that all ionizing photons travel one mean-free-path, hence
\begin{equation}
G(\vec{x},\vec{x}') = \frac{\Theta(|x-x'|-\lambda_{\rm mfp})}{|\vec{x}-\vec{x}'|^2}, 
\end{equation}
where $\Theta$ is the Heaviside step function. This may be rewritten
\begin{equation}
\delta_J(\vec{k}) \propto g(\vec{k}) \delta_k,
\end{equation}    
where $\delta_J(\vec{k})$ and $g$ are the Fourier transforms of the ionizing flux field and filter function respectively. We have
\begin{equation}
g(\vec{k}) \propto \frac{\mbox{Si}(k\lambda_{\rm mfp})}{k},
\end{equation}
where
\begin{equation}
\mbox{Si}(x) = \int_0^x du \frac{\sin(u)}{u}
\end{equation}
and $k=|\vec{k}|$. For small scales $k\lambda_{\rm mfp}\gg1$,  we obtain $\delta_J(k) \propto \delta_k / k$. For large scales we find $\delta_J(k) \propto \delta_k$ as expected from the argument described above.


Thus, we get the following dependence for the function $s$, which can
now be written in terms of $\delta$,
\begin{equation}
s(\delta)=\left[1-K(k)b\delta\right],
\end{equation}
where we have explicitly written in the galaxy bias (noting that it is
the galaxy density rather than the matter density which sources the
ionization field), and defined the function
\begin{equation}
K(k)=K_o\left(1+\frac{k}{k_{\rm mfp}}\right)^{-1},
\end{equation}
where $K_o\propto C$ is a new constant, which from the discussion
following equation~(\ref{fluctuation}) is expected to be smaller than $\sim0.01$. This fitting function interpolates smoothly between the
limiting behaviour on small and large scales. We emphasise that this
formulation is only valid on scales for which fluctuations in the
density field are in the linear regime.

We can now express the fluctuation in
the neutral hydrogen content, noting that this depends both on the
density of galaxies and on the fluctuations in the ionizing
background.
\begin{equation}
\nonumber
\delta_{\rm HI}\approx\left[b\left(1-K(k)\right)-1\right]\delta,
\end{equation}
Similarly, we write the fluctuation in ionized fraction
in terms of the local ionized hydrogen fraction ($x_{i}$) as
\begin{equation}
\delta_{x}\equiv x_{i}/\bar{x}_{i}-1.
\end{equation}
Using the fact that $x_i+x_{\rm HI}=1$, we then find the ionization
fluctuation in terms of the galaxy bias
\begin{equation}
\delta_{x}\approx[{\bar{x}_{\rm HI}}/{(\bar{x}_{\rm HI}-1)}]\,\left[b(1-K(k))-1\right]\delta.
\end{equation}

\subsection{The 21cm PS}
We can now turn to calculating the 21cm PS.  To lowest order in Fourier
space, the velocity fluctuation may be written as,
$\delta_v(\vec{k})=-f\mu^2\delta$ where $\mu$ is the cosine of the angle
between the $\vec{k}$ and line-of-sight unit vectors~(Kaiser~1987),
and\footnote{The quantity $f$ is close to unity at high redshifts, taking
values of 0.974, 0.988 and 0.997 at $z=2.5$, 3.5 and 5.5.}
$f=d\log{\delta}/d\log{(1+z)}$. To leading order in $\delta$, it then
follows from equation~(\ref{Tb}) that the PS of brightness temperature
fluctuations is given by,
\begin{eqnarray}
\nonumber
\label{P21}
P_{\Delta T}&=&\mathcal{T}_{\rm b}^2\left[\left(\bar{x}_{\rm HI}^2P_{\delta\delta} - 2\bar{x}_{\rm HI}(1-\bar{x}_{\rm HI}) P_{\delta x} + (1-\bar{x}_{\rm HI})^2P_{xx}\right) \right.\\
&&\hspace{-10mm}+2f\mu^2\left(\bar{x}_{\rm HI}^2P_{\delta\delta}-\bar{x}_{\rm HI}(1-\bar{x}_{\rm HI})P_{\delta x}\right)+\left.f^2\mu^4\left(\bar{x}_{\rm HI}^2P_{\delta\delta}\right)\right],
\end{eqnarray}
where $\mathcal{T}_{\rm b}=23.8\left[{(1+z)}/{10}\right]^\frac{1}{2}\,$mK,
and $P_{\delta\delta}$, $P_{\delta x}$ and $P_{xx}$ are the PS of density
fluctuations, the PS of ionization fluctuations, and the cross-PS of
ionization and density fluctuations respectively.  Our expression for
$\delta_{x}$ allows the ionization PS and ionization-density cross-PS to be
related to the density PS in a very simple way
\begin{eqnarray}
\nonumber
\label{Pion}
P_{\delta x}&=&-\frac{\bar{x}_{\rm HI}}{1-\bar{x}_{\rm HI}}\left[b(1-K(k))-1\right]P_{\delta\delta}\\
P_{xx}&=&\frac{\bar{x}_{\rm HI}^2}{(1-\bar{x}_{\rm HI})^2}\left[b(1-K(k))-1\right]^2P_{\delta\delta}.
\end{eqnarray} 
In addition, the ionization PS and ionization-density cross-PS are
very simply related to each other
\begin{equation}
P_{xx}=-\frac{\bar{x}_{\rm HI}}{(1-\bar{x}_{\rm HI})}[b(1-K(k))-1]P_{\delta x}.
\end{equation} 

Upon substitution of equation~(\ref{Pion}) into equation~(\ref{P21})
we obtain an expression for the angular dependence of the
post-reionization 21cm PS
\begin{equation}
\label{DLAPS1}
P_{\Delta T}(b)=\mathcal{T}_{\rm b}^2\bar{x}_{\rm HI}^2\left[b(1-K(k))+f\mu^2\right]^2P_{\delta\delta}.
\end{equation}
Thus, when combined equations~(\ref{P21}) and (\ref{Pion}) provide a
simple relation between the density PS and the 21cm PS involving
ionization terms, allowing the full 21cm PS to be utilized in deriving
cosmological constraints. This compares favorably with the
reionization epoch, during which either the $\mu^4$ term alone can be
utilized ~(Barkana \& Loeb~2005), or a functional form for the
ionization PS must be assumed ~(Mao et al.~2008; Rhook et
al.~2008). On large scales fluctuations in the ionization field will
be correlated with density fluctuations. Thus fluctuations in ionizing
background would be degenerate with the galaxy bias $b$. However, on small scales there is scale dependence in the
function $K$ which breaks the degeneracy between ionization and galaxy
bias. In addition, the angular dependence of the PS, which is due to
gravitational infall and hence is independent of galaxy bias or
ionization, breaks the degeneracy between galaxy bias and neutral
fraction.

Equation~(\ref{DLAPS1}) is valid on scales for which fluctuations are
in the linear regime, and we can assume that galaxy bias is not scale
dependent.  In the case where $b=f=1$ and $K_o=0$ (i.e. unbiased
sources at high redshift, with no modulation by the ionizing
background) our expression for the PS reduces to the form of a
uniformly ionized IGM, $P_{\Delta T}(b)=\mathcal{T}_{\rm
b}^2\bar{x}_{\rm H}^2\left[1+\mu^2\right]^2P_{\delta\delta}$ ~(Barkana
\& Loeb~2005).  We note that in the absence of a modulation of power
due to the ionizing background, equation~(\ref{DLAPS1}) could have
been derived directly from considering the galaxy PS (once it is
realized that effect of a line-of-sight velocity is the same for the
optical depth in the 21cm line and the galaxy number density).

\begin{figure*}
\centerline{\epsfxsize=5.9in \epsfbox{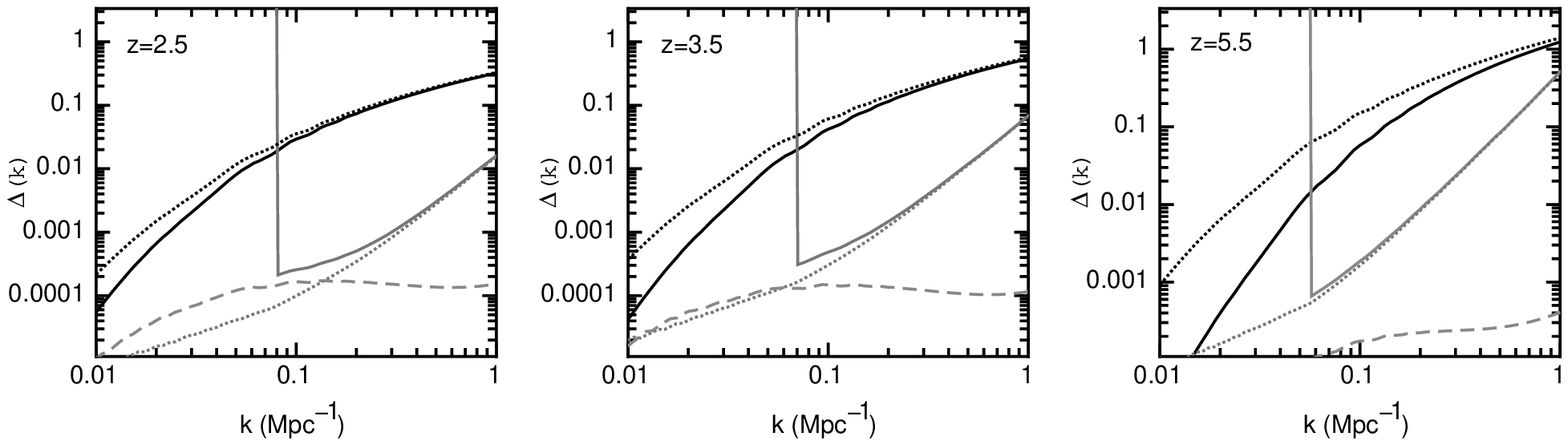}} 
\caption{Examples of the post-reionization 21cm PS at three
redshifts. Two cases are shown, assuming models with $K_o=0$ (dotted
curves) and $K_o=1$ (solid curves). A DLA host mass of $M_{\langle
b\rangle}=10^{11}M_\odot$ was assumed at each redshift. The light
solid curve shows the spherically averaged noise for the MWA5000,
assuming 1000hr integration on each of three fields. The sharp upturn
at low $k$ is due to the assumption that foreground removal prevents
measurement of the PS at scales corresponding to a bandpass larger
than 8MHz. The thermal noise and cosmic variance (including the
Poisson noise) components are plotted as the light dotted and light
dashed curves respectively. The sensitivity curves are plotted within
bins of width $\Delta k=k/10$.}
\label{fig1}
\end{figure*}

Equation~(\ref{DLAPS1}) is valid for a single halo mass. However the DLAs
reside within halos having a range of masses. To account for this we define
an unknown HI mass weighted probability density function
$p(b)db$ for the galaxy bias. By substituting $\delta_{x}$ into
equation~(\ref{Tb}), and integrating over $p(b)$ it is easy to show that
the PS for an arbitrary distribution of bias is
\begin{equation}
\label{meanPS}
P_{\Delta T} = \mathcal{T}_{\rm b}^2\bar{x}_{\rm HI}^2P_{\delta\delta}\left[\langle b\rangle(1-K(k)) +f\mu^2\right]^2,
\end{equation}
where $\langle b\rangle$ is the mean of the bias distribution $p(b)$.
For completeness we also calculate the spherically averaged PS
\begin{eqnarray}
\nonumber
P^{\rm sph}_{\Delta T} &=& \mathcal{T}_{\rm b}^2\bar{x}_{\rm HI}^2P_{\delta\delta}\\
&&\hspace{-7mm}\times\left[\langle b\rangle^2(1-K(k))^2 + \frac{2}{3}f\langle b\rangle(1-K(k)) + \frac{1}{5}f^2\right],
\end{eqnarray}
which, in the case of $\langle b\rangle=1$, $f=1$ and $K_o=0$ reduces
to the form for a uniformly ionized high redshift IGM, $P^{\rm
sph}_{\Delta T}=1.87\times\mathcal{T}_{\rm b}^2\bar{x}_{\rm
HI}^2P_{\delta\delta}$ ~(Barkana \& Loeb~2005).  Note that although
the relation between neutral hydrogen mass and host halo mass is
unknown, the PS depends only on the mean of $p(b)$.  For later use we
define $M_{\langle b\rangle}$ to be the halo mass corresponding to
$\langle b\rangle$.

In Figure~\ref{fig1} we show examples of the dimensionless 21cm PS
after reionization [$\Delta=k^3P_{\Delta T}/(2\pi^2)$]. We assume
$M_{\langle b\rangle}=10^{11}M_\odot$, use the Sheth-Tormen fitting
function for the bias~(Sheth, Mo \& Tormen~2001), and consider three
redshifts, $z=2.5$, $z=3.5$ and $z=5.5$, at which the mean-free-path
for ionizing photons is taken to be $\lambda_{\rm mfp}=600$, 300 and
100 co-moving Mpc, respectively (Bolton \& Haehnelt~2007;
Faucher-Giguere et al.~2008). There is at least a factor of 2
uncertainty in the value of the mean-free-path. However our results
are insensitive to the exact value because the mean-free-path is
larger than the scales accessible to 21cm PS observations (owing to
the requirements of foreground removal) at all redshifts except those
closest to reionization. 

Two models are shown to illustrate the effect of the ionizing
background on 21cm fluctuations. In the first we assume a fiducial
model for which the ionizing background has no effect on the
fluctuations in neutral fraction ($K_o=0$). This is shown by the
dotted curves in Figure~\ref{fig1}. We also show examples of a model
in which $K_0=1$, a value which is unphysically large but is chosen so
as to clearly illustrate the qualitative effect of fluctuations in the
ionizing background on the shape of the 21cm PS. In this model, the
fluctuations in neutral fraction due to the ionizing background are as
strong as (and hence cancel) the effect of the galaxy bias on large
scales. Figure~\ref{fig1} illustrates that the fractional modulation
of the 21cm PS is of order $K_0$, while as discussed in
\S~\ref{thick}, physically plausible values of $K_0$ are of order
$10^{-2}$. Since the fractional modification of the PS is of order
$2K_0$, and our results indicate a fractional change in amplitude
which is smaller than a factor of 2 at $k\sim0.1$ over the redshift
range considered, we therefore find that fluctuations in the ionizing
background should modify the 21cm PS after reionization by $\la 1\%$
on the scales and redshifts of interest.

\subsection{Discussion}
\label{discussion}

The effect of the ionizing background on the 21cm PS can be understood
qualitatively as follows. Firstly, we have argued that the ionizing
background follows the galaxy density field on large scales. As a
result, on large scales regions of overdensity (underdensity) will
have a higher (lower) than average ionizing background. In our linear
formulation, the effect of the ionizing background is degenerate with
galaxy bias on large scales, because in the case of $K_0>0$, it lowers
(raises) the contribution to the 21cm intensity from galaxies with a
particular halo mass. This modification of the 21cm intensity of
galaxies reduces the amplitude of the PS, at all scales $k\ll k_{\rm
  mfp}$. We emphasise that because the fluctuations in the ionizing
background are strongly correlated with density on large scales, they
suppress rather than add additional power to the 21cm PS. The
suppression becomes greater for smaller values of $k/k_{\rm mfp}$. Thus
at fixed scale the effect of a fluctuating background is larger at
higher redshift where the mean-free-path is smaller.

On small scales $k\gg k_{\rm mfp}$, only some fraction of the ionizing
BG is produced locally within the fluctuation. The remainder of the
ionizing background is generated within a region which averages over
many fluctuations, and so has a mean value that is near that of the
background.  As a result, the suppression of power described above,
owing to a correlation between the ionizing background and
overdensity, is not as strong on small scales. Our derivation yields
the variation of the suppression with scale ($\propto k^{-1}$). It
should be noted that our formulation ignores the possible Poisson
contribution to fluctuations in the ionizing background due to
quasars.  Poisson fluctuations introduce additional power into the
21cm PS beyond the component associated with the underlying density
field of galaxies.  The effect of Poisson fluctuations could become
important at low redshift ($z\la3$), where quasars contribute
significantly to the ionizing background. We do not consider the
possible Poisson contribution in the remainder of this paper which
focuses on redshifts above the peak of quasar activity ($z\ga2.5$).

We emphasise that our derivation of the 21cm PS after reionization is
only valid on scales where the density fluctuations are in the linear
regime. It should therefore be noted that there are potential
complications that arise due to non-linearities in the mass and
velocity fields on small scales. Analysis of the galaxy PS derived
from N-body simulations has shown~(Seo \& Eisenstein~2005) that the PS
can be treated as linear on scales greater than 15 co-moving Mpc
(i.e. $k_{\rm max}\lsim0.4$Mpc$^{-1}$) at $z=3.5$, increasing towards
higher redshifts. However, weak oscillatory features in the PS, such as
the baryonic acoustic oscillations, are suppressed on even larger
scales because matter moves across distances on the order of $\sim
5$--10Mpc over a Hubble time\footnote{This characteristic scale of
displacement follows from the fact that $\sigma_8$, the normalization
of the power-spectrum on 8$h^{-1}$Mpc, is of order unity at the
present time.}. As groups of galaxies form, the linear-theory
prediction for the location of each galaxy becomes uncertain, and as a
result noise is added to the correlation among galaxies and hence to
the measurement of the mass PS. The noise associated with the movement
of galaxies smears out the acoustic peak in the correlation function
of galaxies in real space (Eisenstein, Seo \& White~2007; Seo, Siegel,
Eisenstein \& White~2008). The associated reduction of power in the
baryonic acoustic oscillations is found to be in excess of 70\% on
scales smaller than $k_{\rm max}\sim0.4$Mpc$^{-1}$ at $z=3.5$,
corresponding to a length scale of $\sim \pi/(2k_{\rm max})= 3.9$
comoving Mpc (Seo et al.~2008).

More importantly for experiments that aim to measure the PS shape is
the non-linear correction to the PS due to virialised groups of DLAs
(e.g., Tinker et al.~2006; Tinker~2007). The group velocity
dispersion, $\sigma_{\rm vir}$, introduces the so-called
``finger-of-god'' in redshift space which is unrelated to the peculiar
velocity according to linear theory. We can estimate the wavenumber at
which this effect becomes important relative to the Hubble flow as,
$k_{\rm vir}\sim \pi H(z) /(2\sigma_{\rm vir})$. At $z\sim3.5$ we find
$k_{\rm vir}\sim 10$Mpc$^{-1}$.  In the future, an improvement to our
analysis could be made by including the analytic model for the
redshift space galaxy two-point correlation function described by
Tinker~(2007). This model is constructed within the framework of the
Halo Occupation Distribution, and quantifies galaxy bias on
linear and non-linear scales. In addition, the model describes
redshift-space distortions and clustering on both linear and
non-linear scales.  Finally, other non-linear effects may arise that
are not present in galaxy redshift surveys, owing to internal rotation
curve of the neutral hydrogen. However, the corresponding wave numbers
are very large, $k\gg10$Mpc$^{-1}$.

\subsection{sensitivity to the 21cm PS after reionization}
\label{noise}

Before proceeding to discuss the cosmological potential of the
post-reionization 21cm PS, we compute the sensitivity with which it
could be detected.

To compute the sensitivity $\Delta P_{\Delta T}(k,\mu)$ of a
radio-interferometer to the 21cm PS, we follow the procedure outlined
by ~McQuinn et al.~(2006) and Bowman, Morales \& Hewitt~(2007) [see
also ~Wyithe, Loeb \& Geil~(2008)]. The important issues are discussed
below, but the reader is referred to these papers for further
details. The uncertainty comprises of components due to the thermal
noise, and due to sample variance within the finite volume of the
observations.  We also include a Poisson component due to the finite
sampling of each mode~(Wyithe~2008), since the post-reionization 21cm
PS is generated by discrete clumps rather than a diffuse IGM. However
we find that Poisson noise dominates only when $M_{\langle
  b\rangle}>10^{11}M_\odot$ (see Wyithe~2008).  We assume that
foregrounds can be removed over $8$MHz bins, within a bandpass of
$32$MHz ~(McQuinn et al.~2006) [foreground removal therefore imposes a
minimum on the wave-number accessible of
$k\sim0.07[(1+z)/4.5]^{-1}$Mpc$^{-1}$], and consider a hypothetical
follow-up telescope to the MWA which would have 10 times the total MWA
collecting area (we refer to this as the MWA5000). This telescope is
assumed to have an antenna density distributed as $\rho(r)\propto
r^{-2}$ within a diameter of 2km and a flat density core of radius 80m
(see McQuinn et al.~2006). The antennae design is taken to be
optimized at the redshift of observation (so that the physical
collecting area of the array equals its effective collecting area),
with each of 5000 phased arrays (tiles) consisting of 16
cross-dipoles. An important ingredient is the angular dependence of
the number of modes accessible to the array~(McQuinn et al.~2006). We
assume 3 fields are observed for 1000 hr each. For the experiment
described, the signal-to-noise of the PS will be largest over the
decade of scales around $k\sim0.1$Mpc$^{-1}$ (McQuinn et al.~2006;
Wyithe, Loeb \& Geil~2008). The sensitivity curves (within bins of
$\Delta k=k/10$) are plotted as the solid grey lines in
Figure~\ref{fig1}. We find that the both the 21cm PS, as well as the
effect of a non-zero value of $K_o$ would be easily measured using the
MWA5000.

We note that the unprecedented power of cosmic-variance limited 21cm
surveys for constraining cosmological parameters is made possible by
the wide fields of view for the low-frequency telescopes under
construction, combined with the very large volumes available at high
redshift~(Loeb \& Wyithe~2008; Mao et al.~2008). For example, in units
of the Sloan Digital Sky Survey volume ($\mathcal{V}_{\rm
sloan}\sim5.8\times10^8$Mpc$^3$) the telescope described would observe
$\mathcal{V}/\mathcal{V}_{\rm sloan}\sim4.8$,
$\mathcal{V}/\mathcal{V}_{\rm sloan}\sim7.6$ and
$\mathcal{V}/\mathcal{V}_{\rm sloan}\sim9.8$ per observing field, at
$z=2.5$, $z=3.5$ and $z=5.5$, respectively. These large volumes are
obtained because of the very wide field of view available to an array
with a design like the MWA. The array described would have a primary
beam of $A_\theta\sim2000$ square degrees.

\begin{figure*}
\centerline{\epsfxsize=5.9in \epsfbox{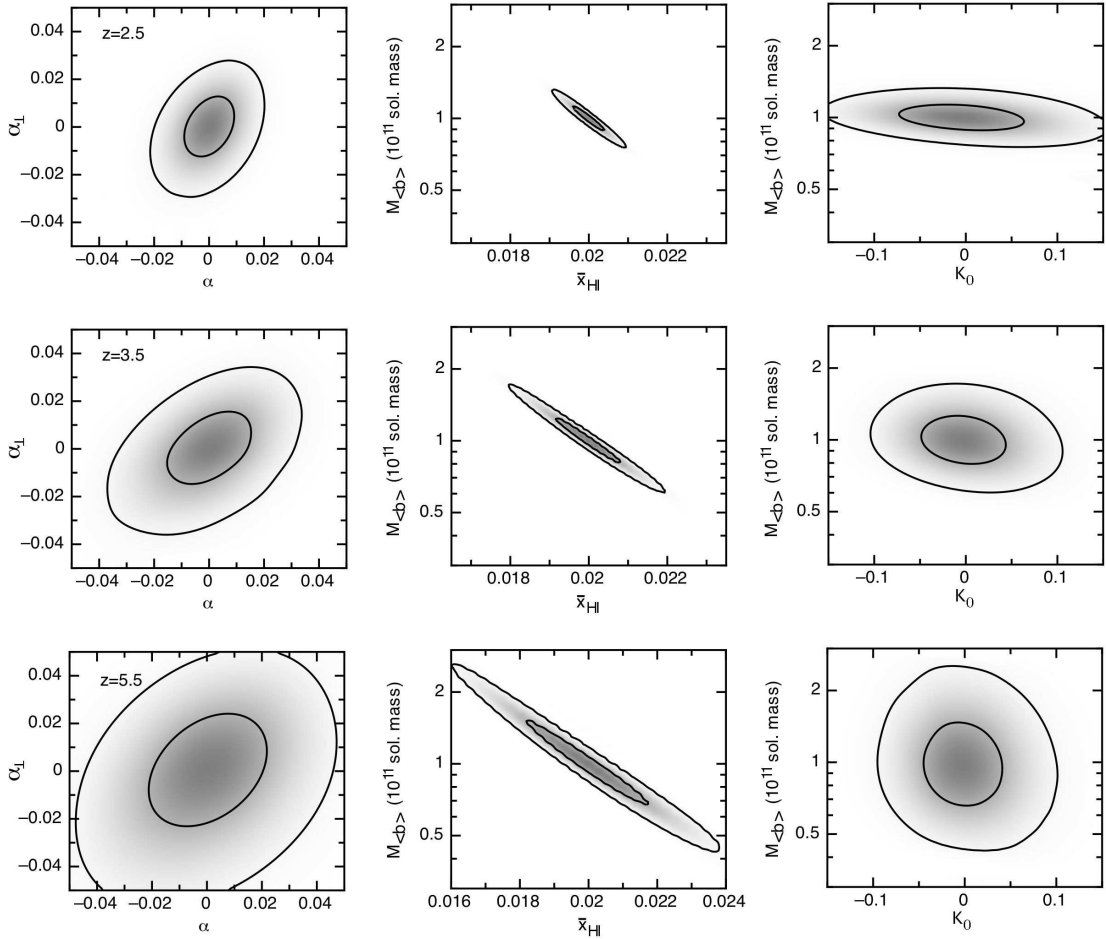}} 
\caption{Constraints on the post-reionization PS distortions
achievable via the AP test at 3 redshifts, assuming a fiducial model
with $K_o=0$. The three columns show parameter sets
$(\alpha,\alpha_\perp)$, $(\bar{x}_{\rm HI},M_{\langle b\rangle})$ and
$(K_o,M_{\langle b\rangle})$. In each case likelihood contours are
shown at 60\% and 7\% of the maximum likelihood. The projections of
these contours onto individual parameter axes correspond to the 68.3\%
and 90\% confidence ranges respectively.}
\label{fig2}
\end{figure*}

\section{Application of the Alcock-Paczynski Test}
\label{secAP}

Like traditional galaxy redshift surveys, the observed 21cm PS is
sensitive to only one underlying PS ($P_{\delta\delta}$), in addition
to the 4 parameters, $\bar{x}_{\rm HI}$ and $\langle b\rangle$,
$k_{\rm mfp}$ and $K_o$. These parameters are related to properties of
DLAs and LLSs, and can be measured through independent means via
quasar absorption line studies (particularly true for $\bar{x}_{\rm
HI}$, $\langle b\rangle$ and $k_{\rm mfp}$).  This situation should be
contrasted with the reionization era where the observed 21cm PS is
also sensitive to long range, non-gravitational fluctuations through
$P_{\delta x}$ and $P_{xx}$.  Provided that the non-linear evolution
of the PS can be properly accounted for, this dominant dependence on
the density PS provides a considerable advantage for the purposes of
cosmological constraints.

To illustrate the potential for cosmological constraints from the
anisotropy of the 21cm PS after reionization, we calculate the
Alcock-Paczynski (AP) effect. Our approach is to specify the general
result of Barkana~(2006) for the distortion of the true PS [$P_{\Delta
T}^{\rm t}(k,\mu)$] resulting from an incorrect choice of
cosmology. This is parameterised in terms of the dilation parameters
$\alpha$ and $\alpha_\perp$, which describe the distortions between
the transverse and line-of-sight scales, and in the overall scale,
respectively. These are defined such that $(1+\alpha)$ is the ratio
between the assumed and true values of ($D_{\rm A}H$), while
$(1+\alpha_\perp)$ is the ratio between the assumed and true values of
the angular diameter distance, $D_{\rm A}$. In the AP test, the
correct cosmology is inferred by finding cosmological parameters for
which $\alpha=\alpha_\perp=0$.

To calculate the AP effect we apply equation~(8) in ~Barkana~(2006) 
\begin{eqnarray}
\nonumber
P_{\Delta T}(k,\mu)&=&(1+\alpha-3\alpha_\perp)P_{\Delta T}^{\rm t}+(\alpha\mu^2-\alpha_\perp)\frac{\partial P_{\Delta T}^{\rm t}}{\partial \ln{k}}\\
&&+\alpha(1-\mu^2)\frac{\partial P_{\Delta T}^{\rm t}}{\partial\ln{\mu}}
\end{eqnarray}
to the PS in equation~(\ref{meanPS}). Here the PS $P_{\Delta T}$ and
$P_{\Delta T}^{\rm t}$ are evaluated at the observed $\vec{k}$. This
procedure results in a modified PS that is related to the true density PS
($P_{\delta\delta}^{\rm t}$) via
\begin{eqnarray}
\label{APPS}
\nonumber
P_{\Delta T}(k,\mu) &=& \mathcal{T}_{\rm b}^2 \bar{x}_{\rm HI}^2  P_{\delta\delta}^{\rm t} \left(\langle b\rangle(1-K(k))+f\mu^2\right)^2\\
\nonumber
&&\hspace{0mm}\times\left[(1+\alpha-3\alpha_\perp) + \frac{d\ln{P_{\delta\delta}^{\rm t}}}{d\ln{k}}(\alpha\mu^2-\alpha_\perp)\right]\\
\nonumber
& -& \mathcal{T}_{\rm b}^2 \bar{x}_{\rm HI}^2  P_{\delta\delta}^{\rm t} (\langle b\rangle(1-K(k))+f\mu^2)\\
\nonumber
&&\hspace{0mm}\times\left[2\frac{k}{k_{\rm mfp}}\frac{-K(k)}{1+\left({k}/{k_{\rm mfp}}\right)}\langle b\rangle(\alpha\mu^2-\alpha_\perp)\right.\\ 
&&\hspace{25mm}-\left.4 \alpha\left(1-\mu^2\right) f\mu^2 \right].
\end{eqnarray}
Barkana~(2006) expanded the general form of this equation to find the
linear combinations of the true PS $P^{\rm t}_{\delta\delta}$, $P^{\rm
t}_{\delta x}$ and $P^{\rm t}_{x x}$ which form the coefficients of terms
containing $\mu^0$, $\mu^2$, $\mu^4$ and $\mu^6$.  His result shows that in
general, the parameter $\alpha$ (which corresponds to anisotropy) mixes
$P^{\rm t}_{\delta\delta}$, $P^{\rm t}_{\delta x}$ and $P^{\rm t}_{x
x}$. As a consequence, the $\mu^6$ term, which arises through the AP
effect, must be isolated in order to probe the anisotropy parameter
$\alpha$. On the other hand, equation~(\ref{APPS}) is sensitive only to
$P^{\rm t}_{\delta\delta}$. As a result, the coefficients of all powers of
$\mu$ in the post-reionization PS can be utilised in the AP effect to find
the cosmological parameters that yield $\alpha=\alpha_\perp=0$.

\subsection{AP constraints on the PS}

We next use equation~(\ref{APPS}) to calculate the permissible region
of parameter space $\vec{p}=(\alpha,\alpha_\perp,\bar{x}_{\rm
HI},M_{\langle b\rangle},K_o)$ around a true solution with PS $P^{\rm
t}_{\delta\delta}$ and $\vec{p}_o= (0,0,0.02,10^{11}M_\odot,0)$.  The
fiducial value of $M_{\langle b\rangle}$ is chosen to lie in the
middle of the mass range for DLAs measured from the cross-correlation
analysis of Cook et al.~(2006). We have assumed that the
mean-free-path of ionizing photons is known a'priori, and do not fit
it as a free parameter. As part of this procedure, we assume an
estimate for the sensitivity of the future low-frequency
interferometer described in \S~\ref{noise} to the 21cm PS [$\Delta
P_{\Delta T}(k,\mu)$], and construct likelihoods
\begin{equation}
\ln{\mathcal{L}(\vec{p})} = -\frac{1}{2}\sum_{k,\mu}\left(\frac{P_{\Delta T}(k,\mu,\vec{p})-P_{\Delta T}^{\rm t}(k,\mu,\vec{p}_o)}{\Delta P_{\Delta T}(k,\mu)}\right)^2,
\end{equation}
where the sum is over bins of $k$ and $\mu$. To estimate the constraints achievable via the AP effect, we modify
the linear power spectrum to include the nonlinear erasure of the acoustic peaks based on the work of Eisenstein et al.~(2006). Specifically,  we use 
\begin{equation}
P^{\rm t}_{\delta\delta}(k) = \left(P_{\delta\delta}(k) - P^{\rm smth}_{\delta\delta}(k)\right) \exp{\left(-k^2\Sigma_{\rm nl}^2/2\right)} + P^{\rm smth}_{\delta\delta}(k),
\end{equation}
where $P^{\rm smth}_{\delta\delta}$ is the ``no wiggle'' form from Eisenstein \& Hu (1999), and $\Sigma_{\rm nl}=3.9[(1+z)/4.5]^{-1}$Mpc in the high redshift limit. In addition, to account for the possibility of non-linearity in the smooth PS at small scales we restrict our fitting to wave
numbers $k_{\rm max}<0.4$Mpc$^{-1}$ (see \S~\ref{discussion}).

The results are shown in Figure~\ref{fig2} for three redshifts,
$z=2.5$, 3.5 and 5.5.  Likelihood contours are shown at each redshift
for the parameter sets $(\alpha,\alpha_\perp)$, $(\bar{x}_{\rm
  HI},M_{\langle b\rangle})$ and $(K_o,M_{\langle b\rangle})$.  For
each set the likelihood is marginalised over the remaining 3
parameters assuming flat prior probabilities. The exception is
$\bar{x}_{\rm HI}\approx 0.02\pm0.002$, which we have assumed to be
known a'priori (with Gaussian errors) from future observations of
quasar absorption spectra. The assumed uncertainty in $\bar{x}_{\rm
  HI}$ corresponds to a factor of $\sim 2$ improvement over existing
measurements~(Wolfe, Gawiser \& Prochaska~2005). Figure~\ref{fig2}
shows that the angular dependence of the PS breaks degeneracy between
the different parameters, allowing $\bar{x}_{\rm HI}$, $M_{\langle
  b\rangle}$, $\alpha$, and $\alpha_\perp$ to each be measured from
the fitting. In addition, the departure from the underlying shape of
the PS allows the value of $K_o$ to also be constrained.

We note that by including the uncertainties in the AP effect (through
$\alpha$ and $\alpha_\perp$), our derived uncertainties for
non-cosmological parameters ($M_{\langle b\rangle}$, $\bar{x}_{\rm
HI}$ and $K_o$) include the uncertainty in the underlying matter power
spectrum.  An exception is that the uncertainty in the PS amplitude
(which is proportional to the normalization of the primordial PS,
$\sigma_8$) is degenerate with $\bar{x}_{\rm HI}$.  However, since the
error in $\bar{x}_{\rm HI}$ is comparable to or larger than the
fractional error in $\sigma_8$, this does not add additional
uncertainty relative to the constraints shown in Figure~\ref{fig2}.
An important additional uncertainty may also be introduced through the
uncertainty in the primordial PS index or through a running spectral
index, which we have not considered in this study.

At $z\la3.5$ the neutral fraction constraints are improved over the
assumed prior information from the IGM. The mass of the DLA hosts is
constrained to high precision (tens of percent, Wyithe~2008), with
more accurate estimates at lower redshifts. In addition, the value of
$K_o$ is constrained to be smaller than a few hundredths (with minimal
redshift dependence), although this in not at a level comparable to
the physically expected value of $\sim10^{-2}$. Thus, on scales larger
than the mean-free-path, the perturbation due to the ionizing
background ($\langle b\rangle K_o\delta$) is constrained at a level
that is smaller than $\langle\delta^2\rangle^{1/2}$. On scales smaller
than the mean free path, this perturbation is suppressed by $(k/k_{\rm
  mfp})^{-1}$. The cosmological constraints follow from the precision
with which the distortion of the 21cm PS (as described by $\alpha$ and
$\alpha_\perp$) can be measured.  We find that distortions owing to
the assumption of an incorrect cosmology could each be constrained at
a percent level.

\begin{figure*}
\centerline{\epsfxsize=5.9in \epsfbox{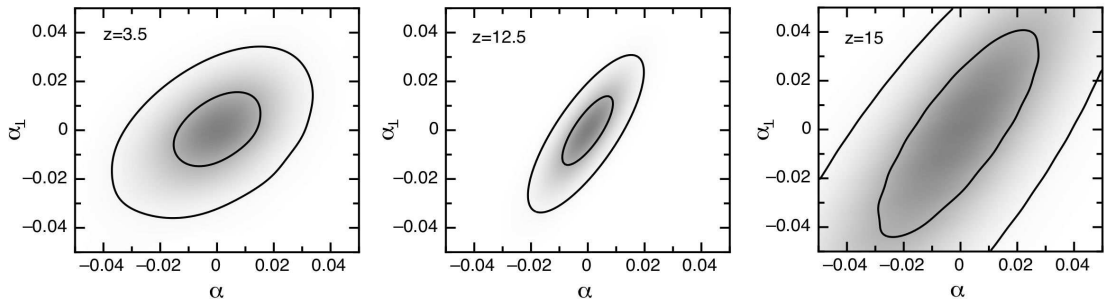}} 
\caption{{\em Central and Right panels:} Constraints on the
pre-reionization PS distortions achievable via the AP test at two
redshifts. The results correspond to an optimistic high-redshift case,
with $\bar{x}_{\rm HI}=1$, and $P_{xx}=P_{\delta x}=0$. The panels
show the parameter set $(\alpha,\alpha_\perp)$. In each case,
likelihood contours are shown at 60\% and 7\% of the maximum
likelihood with each set marginalised over $\bar{x}_{\rm HI}$ (which
we have assumed to be unconstrained a'priori). The projections of
these contours onto individual parameter axes correspond to the 68.3\%
and 90\% confidence ranges respectively. For comparison, ({\em left
panel}) the $K_o=0$ case at $z=3.5$ is repeated from Figure~\ref{fig2}.}
\label{fig3}
\end{figure*}

\subsection{Comparison with high redshift constraints}

Before concluding we show for comparison (Figure~\ref{fig3}) the
constraints on the parameter set $(\alpha,\alpha_\perp)$ for
optimistic high redshift cases ($z=12.5$ and 15), where we assume
$\bar{x}_{\rm HI}=1$ and fluctuations dominated by the density field
($P_{xx}=P_{\delta x}=0$). These cases have fitted parameters
$\vec{p}=(\alpha,\alpha_{\perp},\bar{x}_{\rm HI})$, and correspond to
the IGM prior to the formation of galaxies and ionized bubbles. The
smaller $\bar{x}_{\rm HI}$ at $z<6$ relative to the pre-reionization
epoch is offset by the galaxy bias factor enhancement as well as the
growth of structure, and by lower foreground contamination. With
respect to constraining cosmological parameters, we find that the
post-reionization 21cm PS would be competitive with the most
optimistic expectations for the 21cm PS at high redshift.

\section{Conclusions}

In this paper we have derived the 21cm power spectrum (PS) following
the completion of reionization. Our approach is to derive the PS
within the formalism that has been developed to calculate the 21cm PS
during reionization~(Barkana \& Loeb~2005; McQuinn et al.~2006; Mao et
al.~2008). By framing the derivation this way we are able to directly
compare the relative merits of cosmological constraints from 21cm
surveys at redshifts prior to and post reionization. We have derived
expressions for the components of the post-reionization 21cm PS that
are due to the density field, the ionization field, and their
cross-correlation. As is the case prior to reionization, we find that
these components contribute to the observed PS in proportions that
depend on the angle relative to the line-of-sight along which the
power is measured. However, in difference from the situation prior to
reionization, we have shown that all components of the 21cm PS are
directly proportional to the PS of the underlying matter fluctuations,
with a small but predictable modulation on scales below the
mean-free-path of ionizing photons. We have derived the form of this
modulation, and have shown that its effect on the observed PS will be
at less than the 1\% level for physically plausible astrophysical
models of DLA systems. The simplicity of the 21cm PS after
reionization stands in contrast to the astrophysical uncertainty
during the reionization epoch, where HII regions dominate the 21cm
signal. This simplicity will enhance the utility of the 21cm PS after
reionization as a cosmological probe (Loeb \& Wyithe 2008) by removing
the need to separate the PS into physical and astrophysical components
(Barkana \& Loeb 2005).

To illustrate the utility of the 21cm PS after reionization as a
cosmological probe, we have examined the Alcock-Paczynski~(1979)
test. Our calculations show that the next generation of low-frequency
arrays could measure the angular distortion of the PS to around $\sim
1\%$ at $z\sim3.5$. It has previously been shown that the scale of
Baryonic Acoustic Oscillations, which constitutes a standard ruler
~(Blake \& Glazebrook~2003; Seo \& Eisenstein~2005; Eisenstein et
al.~2005; Padmanabhan et al.~2007), can be also be used to
independently probe $H$ and $D_{\rm A}$ in 21cm PS, and hence to
measure the equation of state of the dark energy ~(Wyithe, Loeb \&
Geil~2008; Chang et al.~2007). With the caveat that non-linear
evolution of the 21cm PS must be quantitatively understood, our simple
analysis indicates that the precision achievable via the
Alcock-Paczynski~(1979) test using the 21cm PS after reionization
could be better than those available via a 21cm fluctuation
measurement of the acoustic scale of baryonic oscillations for a given
observing strategy (Shoji, Jeong \& Komatsu 2009).
More generally, our analysis shows that the 21cm PS after reionization
shares the same favorable features as a galaxy redshift survey. The
advantage of using 21cm fluctuations lies in the fact that individual
sources need not be resolved. This would allow PS measurements using
21cm fluctuations to be extended to higher redshifts.

\bigskip

\noindent
{\bf Acknowledgments.} The research was supported by the Australian
Research Council (JSBW), by NASA grants NNX08AL43G and LA, by FQXi, and
by Harvard University funds (AL). We thank an anonymous referee for correcting an earlier error.

\label{lastpage}
\end{document}